\documentclass[a4paper]{article}

\newcommand{\be}{\begin{equation}}
\newcommand{\ee}{\end{equation}}
\newcommand{\ba}{\begin{eqnarray}}
\newcommand{\ea}{\end{eqnarray}}

\begin{document}
\noindent{\Large  \bf Instanton and Superconductivity in  Supersymmetric  CP(N-1) Model}
\vskip .2 in
\newlength{\abswidth}
\setlength{\abswidth}{0.85\linewidth}
\hfill{\begin{minipage}{\abswidth}
{\bf Shinobu Hikami  and Takuro Yoshimoto}
\vskip 1mm
{\footnotesize Department of Basic Sciences,
 University of Tokyo.\\
Meguro-ku, Komaba, Tokyo 153-8902, Japan.
\\
E-mail:hikami@dice.c.u-tokyo.ac.jp and yoshimoto@dice.c.u-tokyo.ac.jp}
\vskip 1mm
{\bf Abstract}\\
 The two dimensional supersymmetric CP(N-1) model  has a striking similarity to the 
${\mathcal{N}}=2$ supersymmetric gauge theory in  four dimensions. 
The BPS mass formula and the curve of the marginal stability (CMS),
which  exist in the four dimensional gauge theory,
appear in this two dimensional  CP(N-1) model. 
These two quantities are 
 derived by a one-dimensional n-vector spin model
in the large n limit for the $N=2$ case.
This mapping is further investigated at the critical point.
 An application of the study of the BPS mass formula is proposed to the 
phenomena of the spin and charge separations in the Higgs phase.
\end{minipage}}
\vskip 4mm 
This article is the extension of the bosonic CP(N-1) model, for which the relation to the
superconductivity has been studied \cite{Hikami}. 
The supersymmetric CP(N-1) model has been applied to the two dimensional 
fluctuation phenomena which are related to the Higgs mechanism \cite{Dadda2}. There are several examples
in the condensed matter physics,
which have both spin and charge fluctuations, in the strongly correlated
systems. For instance, the high temperature superconductor
in  the underdoped region may be one of examples of the systems, where the 
magnetic spin
fluctuations and the gauge field fluctuations become important.
Although our analysis is restricted to the  phenomenological one, we
 intend to study universal
behaviors of the fluctuations of the bosonic and fermionic
excitations based on the supersymmetric model.
 
The two dimensional supersymmetric CP(N-1) model has  
 ${\mathcal{N}}=2$ supersymmetry and 
an axial anomaly \cite{Witten1}.
A striking similarity between the four dimensional supersymmetric QCD 
 and the two
dimensional supersymmetric CP(N-1) model exists \cite{Dorey}.
The electric  charge and the topological charge give  the BPS mass spectra
in the four dimensional supersymmetric QCD \cite{Seiberg}.
In the four dimension, a weak coupling region is separated from a
strong coupling region by a curve of marginal stability (CMS), 
where the bound state
becomes marginal and BPS masses can decay. 
This CMS has also a  correspondence in the CP(N-1) model
 \cite{Dorey,Shifman}.

In the four dimensional ${\mathcal{N}}=2$ gauge theory, the partition function is given by
the statistical sum over random partitions,  which leads to 
the prepotential and the spectral curve \cite{Nekrasov}.
In the two dimensional supersymmetric CP(N-1) case, it is  desirable to discuss 
such a partition function from a useful representation. 
We find a simple model which gives the same effective potential and the curve of 
marginal stability (CMS) as the two dimensional supersymmetric CP(1) model.
This model is an exactly solvable n-vector spin model,
which was studied for the large order behavior of the $\frac{1}{n}$ expansion
through the instanton analysis \cite{HB}.
 We reinterprete this n-vector  
model by the analytic continuation of the temperature T, like the 
Lee-Yang zeroes of the Ising model, in the complex temperature plane.

The supersymmetric  CP(N-1) model  is described by the Lagrangian,
\ba
{\mathcal{L}} &=& \frac{1}{ 2 g}[(D_\mu Z_i)^{\dagger}(D_\mu Z_i) - \lambda (Z_i^{*} Z_i-1)]\nonumber\\
&& + \frac{1}{ 2 g}[ i \bar \psi^i  \gamma_\mu D_\mu \psi_i + \frac{1}{2}(\sigma^2 +
\pi^2)- \sqrt{\frac{1}{2}}\bar\psi^i(\sigma + i \pi \gamma_5)\psi_i],
\ea
where $\psi_i$ is two component fermion. The fields $\lambda$, $\sigma$ and $\pi$ are
auxiliary fields. 
There appears a dynamically generated mass $m$, which is same for the boson and for the fermion
part by the supersymmetry, and it is evaluated by the renormalization group 
$\beta$ function in a one-loop order as
\be\label{mL}
\Lambda= \mu e^{1 - \frac{\pi}{g}}.
\ee
We replace the renormalization point $\mu$ by the twisted mass $m$, by putting 
\be
m= \mu.
\ee
The dynamically generated mass $\Lambda$ is small when the coupling $g$ is small.
In the four dimensional SQCD,
the massive excitations of monopole, dyon, and Noether charges are evaluated
by  the elliptic integrals. The monodoromy of the three
singularities at $x=\pm 1$ and $x=u$ determine these mass formula, where $u$ is a moduli parameter, 
representing the value 
of the Higgs field. 
In the two dimensional $CP(1)$ case, 
we have a different monodoromy with a single parameter $u= m_1^2 = m_2^2 $.
The electric charge $n_e$ and topological charge $n_t$, which are integers, are combined with
the mass $m$ and the dual mass $m_D$ to form the charge $Z_{n_e,n_t}$ as
\be\label{ZqT}
Z_{n_e,n_t} =  m n_e +  m_D n_t
\ee 
In general, $m$ and $m_D$ are 
complex numbers, and if $m$ and $m_D$ have same phase, i.e. $m/m_D$ is real number,
the bound states of $(n_e,n_t)$ become marginal. This marginal case
is represented by a curve in the complex $m^2$ plane, and called as the curve of marginal
stability (CMS), which separates the weak coupling region and the strong coupling region.
Following the argument of the twisted theory, the vacuum angle $\theta_{eff}$ and the coupling constant
$g_{eff}$, which are modified by the quantum corrections from the bare values, are combined as
\be\label{tau}
\tau = -  \frac{i}{g_{eff}} + \frac{\theta_{eff}}{2\pi}, 
\ee
and the relation to the $\tilde \Lambda$, which is modified by the $\theta_{eff}$ value, is
\be\label{thetapi}
\frac{\tilde \Lambda}{m} = \frac{1}{2} e^{1 - i \pi \tau} =
\frac{1}{2}e^{1 - \frac{\pi}{g_{eff}} -  \frac{i}{2} \theta_{eff}}.
\ee
There is a critical point at $g_{eff}=\theta_{eff}= \pi$, which reads
$4 (\frac{\tilde \Lambda}{m})^2 = -1$.
The twisted chiral superfield becomes
\be
\Sigma = \sigma + \sqrt{2}\vartheta^\alpha \tilde \chi_\alpha + \vartheta^\alpha \vartheta_\alpha
S.
\ee
The effective  twisted superpotential $\tilde W$ with twisted masses $m_i$ \cite{Hanany} is  
obtained by this twisted chiral superfield $\Sigma$.
The condition $\partial \tilde W/\partial \Sigma = 0 $ implies 
\be
\prod_{i=1}^N (\sigma + m_i) - \tilde \Lambda^N = \prod_{i=1}^N (\sigma - e_i) = 0.
\ee
For the simplicity, from now on we consider $N=2$ case.
The supersymmetric vacuum state is given by $\sigma = e_1,e_2$
according to the two different boundary condition at the infinity. Thus we get
\be\label{Zkl}
Z_{12} = 2 [\tilde W(e_1) - \tilde W(e_2)]\nonumber\\
= \frac{1}{2\pi} [ N (e_1 - e_2) - \sum_{i=1}^2 m_i {\rm ln} (\frac{e_1 + m_i}{e_2 + m_i})].
\ee
By putting $m_1 = - m_2 = - m/2$, we have
$\sigma^2 - \frac{m^2}{4} = \tilde \Lambda^2$.
From (\ref{Zkl}), $m_D$ is expressed by
\be\label{mD}
m_D =  \frac{i}{\pi} [\sqrt{m^2 + 4 \tilde \Lambda^2} + \frac{m}{2}
{\rm ln} (\frac{m-\sqrt{m^2 + 4 \tilde \Lambda^2}}{m+ \sqrt{m^2 + 4 \tilde \Lambda^2}})]
\ee
since $\sigma = \pm \sqrt{m^2/4 + \tilde\Lambda^2}$.

In  the strong coupling region $|m| \ll  |\tilde \Lambda|$,
 BPS states becomes  only $(n_e=0,n_t=1)$ and $(n_e=1,n_t=-1)$, and $|m_D|$ becomes
larger than $|m|$.
In the weak coupling region $ |\tilde \Lambda| \ll |m|$, these two states are bounded, and
other $(n_e,n_t)$ BPS states appear, and $m_D$ can be expanded in the power of $\tilde \Lambda/m$,
in which the whole instanton contributions appear.
There is a boundary,  called as the curve of the marginal stability
(CMS) in the complex mass parameter $m^2$, where the restructuring of the 
BPS states occurs. On this CMS, the masses of dyons and solitons become same as the elementary
mass $m$.
The curve of marginal stability is expressed as the following equation \cite{Shifman}
\be\label{CMS}
{\rm Re} [ {\rm ln}\frac{1+ \sqrt{1 + 4\tilde \Lambda^2/m^2}}{1 - \sqrt{1 + 4 \tilde \Lambda^2/m^2}}
- 2 \sqrt{1 + 4 \tilde \Lambda^2 /m^2}] = 0.
\ee
For the comlex $m^2$ value at a fixed $\tilde \Lambda$, the solution of above equation gives
the curve of marginal stability CMS, which devides the complex $m^2$ plane into two regions,
the weak coupling and strong coupling regions. 
There is a singular point on this CMS, which is  the point  $ 4\frac{ \tilde \Lambda^2}{m^2}=-1$,
and
the value of $m_D$ in (\ref{mD}) becomes vanishing. This critical point
is realized for $g_{eff}=\theta_{eff}=\pi$ as shown in (\ref{thetapi}).
When $4 \frac{\tilde \Lambda^2}{m^2}$ is a positive real number, the solution of the equation
for CMS is $4 \frac{\tilde \Lambda^2}{m^2} = (0.663)^2 = 0.440$.
 
We now discuss the one dimensional n-vector  model, which has an instanton in the large n limit.
This n-vector  model has been studied for the large order behavior of the $\frac{1}{n}$ expansion
in \cite{HB}. The large order behavior is governed by the instanton.
The Hamiltonian $\mathcal{H}$ of this model is
\be
{\mathcal{H}} = - J \sum_{i=1}^{M-1} \vec S_i\cdot \vec S_{i+1},
\ee
with a condition,
\be
|\vec S_i|^2 = \sum_{m=1}^n S_i^2(m) = n.
\ee
The partition function $Z$  for this model 
is obtained easily by the integration of the angles between the neighboring spins,
\be
Z = [ (\frac{n J}{2 k T})^{1-\frac{n}{2}}\Gamma(\frac{n}{2}) I_{\frac{n}{2}-1}(\frac{nJ}{kT})]^{M-1}
\ee
where $I_\nu (z)$ is a modified Bessel function.
We use the parameters $\nu = \frac{n}{2}$ and $Y = \frac{2J}{kT}$ for convenience.
This modified Bessel function has an integral representation,
\be\label{coeff}
I_{\nu}(\nu Y) = \frac{1}{\sqrt{\pi} \Gamma(\nu + \frac{1}{2})} 
(\frac{\nu Y}{2})^\nu 4^\nu e^{-\nu Y}
\int_0^\infty e^{-F(t)}dt,
\ee
with
\be\label{EF}
F(t) = (\nu + \frac{1}{2})t - (\nu - \frac{1}{2}){\rm ln}(1 - e^{-t}) - 2 \nu Y e^{-t}.
\ee
In the large $\nu$ limit, the saddle point equation becomes 
$\partial F(t)/\partial t = 0$.
The two saddle points $t_{+}$ and $t_{-}$ are
\be
e^{-t_{\pm}} = \frac{Y -1 \pm \sqrt{1 + Y^2}}{2 Y}.
\ee
Using these values, we find the values of the exponent $F$,
\be
F(t_{\pm}) = \nu \left( 1 - Y \mp \sqrt{1 + Y^2} - {\rm ln}\frac{-1 \pm \sqrt{1+Y^2}}{2 Y^2}\right).
\ee
The difference becomes
\be\label{FAB}
F(t_{-}) - F(t_{+})  
=2 \nu \left(
\frac{1}{2} {\rm ln}\frac{1 - \sqrt{1 + Y^2}}{1 + \sqrt{1 + Y^2}} + \sqrt{1 + Y^2}\right).
\ee
Above quantity is known to give the dominant contribution to the large order behavior
\cite{HB}. In the $\frac{1}{n}$ expansion, we obtain
\be
I_\nu(\nu Y) = \frac{1}{2 \pi \nu}\frac{e^{\nu \eta}}{(1 + Y^2)^{\frac{1}{4}}}( 1
+ \sum_{k=1}^\infty \frac{u_k(t)}{\nu^k}),
\ee
where we obtain $\eta$, by collecting the coefficients in (\ref{coeff}),
\be
\eta= \sqrt{1 + Y^2} + \frac{1}{2}{\rm ln}\frac{ \sqrt{1 + Y^2} -1}{ \sqrt{1 + Y^2} + 1}.
\ee
This leading term $\nu \eta$ is different from (\ref{FAB}) only by $\nu \pi i$.
The term $u_k(t)$ is  determined by the recursion equation \cite{Abramowitz}. The parameter $t$ 
is $(1 + Y^2)^{-1/2}$.
 It is an asymptotic expansion, which glows like $(k-1)!/[F(t_{+})-F(t_{-})]^k$.
If we identify $\frac{J}{kT}= \frac{\tilde \Lambda}{m}$, ($Y^2= \frac{4\tilde \Lambda^2}{m^2}$),
we find  the expression for $m_D/m$ in (\ref{FAB}) except a factor $i/\pi$.  
Expanding the logarithmic term in $F(t_{+})-F(t_{-})$ by $\sqrt{1 + Y^2}$, and taking a double scaling
limit of
 $Y^2 \rightarrow -1$ and $N \rightarrow \infty$, we obtain the double scaling relation, 
\be\label{double}
F(t_{+})-F(t_{-}) \sim \frac{1}{3} n (1 + Y^2)^{\frac{3}{2}}.
\ee
This double scaling limit corresponds to the superconformal point $\frac{4\tilde\Lambda^2}{m^2}= -1$.
Note that other models may exist which give the same double scaling limit, for instance $\lambda \phi^4$ model,
but other models do not give the same CMS. The n-vector model is expressed by the Bessel function, which
is a confluent hypergeometrical function, and the monodromy of this function is important. This is contrasted
with the case of the four dimensional $\mathcal{N}=2$ supersymmetric gauge theory, where
the periods of the spectral curve are given by the elliptic functions.

Since the correspondence of $J/kT = \tilde \Lambda/m$ requires $Y=2J/kT= \pm i$ for the superconformal
point, we investigate the correlation length of this n-vector model in the large n limit.
 One dimensional n-vector model is solved by the transfer matrix method.
The transfer matrix is 
\be
T = e^{\frac{nJ}{kT} \vec S\cdot \vec S^\prime}.
\ee
The eigenvalues are given by \cite{Balian}
\be
T_l = C I_{\frac{n}{2}- 1 + l}(\frac{nJ}{kT}),
\ee
where $l=0,1,2,...$, and $C$ is a constant.
The spin-spin correlation function for the distance $r$ is given by
\be
<\vec S(0)\cdot \vec S(r)> = \left(\frac{T_1}{T_0}\right)^r = 
\left(\frac{I_{\nu}(\nu Y)}{I_{\nu-1}(\nu Y)}\right)^r,
\ee
with $\nu = \frac{n}{2}$.
If $Y$ is a real, we have a finite correlation length $\xi$ since 
$I_{\nu}(\nu Y) < I_{\nu -1}(\nu Y)$, and there is no phase transition at finite temperature.
The correlation length $\xi$ is given by
\be
\xi^{-1} = {\rm ln}(\frac{I_{\nu-1}(\nu Y)}{I_{\nu}(\nu Y)}).
\ee
This quantity becomes positive for the real $Y$, however, when $Y$ is a pure imaginary number $Y=-i|Y|$, 
there appears a phase transition
with the infinite correlation length by the degeneracy of the eigenvalues of the transfer
matrix.
When $Y$ is imaginary, the modified Bessel function is expressed by the Bessel function $J$ as
 $I_{\nu}(\nu Y ) = e^{-\nu \pi i/2}J_{\nu}( \nu |Y|)$.
We find in the large $\nu$ limit, there appears  a crossing at $Y=\pm i$, 
\be
J_{\nu}(\nu |Y|) = J_{\nu-1}(\nu |Y|).
\ee

The point $Y = \pm i$ is the critical point, which corresponds to $4\frac{\tilde \Lambda^2}{m^2}
= -1$.
Since the Bessel function $J_{\nu}(z)$ is an oscillating function, there appear successive
degeneracies for $\nu$ ($\nu = \frac{n}{2} -1 + l$). 
The succesive transitions due to the degeneracy of
the angular quantum numbers $l=0,1,2,...$, which represent s,p,d,f,....,states.
 Such successive transitions give a cut in the large n limit beyond $|Y|>1$, which is a low
temperature phase. The mass of the inverse of the correlation length $\xi$, which is finite in the 
high temperature region, becomes zero below the transition temperature $T_c$.
It will be interesting to note that such phase transitions also appear for the one dimensional 
n-vector model, with a real positive
$\frac{J}{kT}$, for $n <1$, as shown previously in \cite{Balian}. 
At the critical point, where the cut appears in the square root singularity in (\ref{CMS}), 
the model may be relevant to the massless Thirring model \cite{HM}.

We now briefly discuss the relation to the Higgs phase. 
The CP(N-1) model has been discussed as an equivalent model of the N component scalar QED model, 
or Ginzburg Landau model
with a gauge field \cite{Hikami}, for the critical behavior near the transition temperature.
The later is a model for the superconductor, and by the renormalization group analysis in the large N, 
it has been shown 
that two model 
is equivalent when there is a stable fixed point, for instance the critical exponents become 
same. The superconductor corresponds to N=1 case in the 
N-component scalar QED model. In this article, we have discussed a supersymmetric CP(N-1) model.
We have introduced twisted masses, which give the anisotropy for the order parameters. Under this anisotropy,
the soliton and dyon (charged soliton) represent the kink singularities, and the instanton becomes a bound state
of two opposite kinks. This is similar to the solution of the anisotropic two dimensional $S^2$ instanton, which is 
made of two vortices (merons) \cite{HT}. In two dimesions, there is no long range order, but there is 
a Kosteritz-Thouless phase. The correlation in the long distance has a behavior of the algebraic decay, and
this corresponds to a vortex bound sate. 

The dynamical mass generation of the supersymmetric CP(N-1) model is interpreted
as a formation of a gap. In the four Fermi interaction model, this mass generation is intepreted as 
a superconductor gap although there is no true long range order since the dimension is two. 
In the supersymmetric CP(1) model with a twisted mass, there is a dynamical generated mass $\tilde \Lambda$,
which represents a gap. In the weak coupling case, this gap is made of the bound state of the
soliton (0,1) and dyon (1,-1).
It is quite interesting to note that similar bound state is suggested experimentally as a pseudo gap in the 
high temperature superconductors. The magnitude of the pseudo gap is same order as the superconducting gap
\cite{Timusk}. 
If the supersymmetric CP(1) model deformed by a twisted mass is relevant to the high temperature superconductivity phase 
diagram, the curve of marginal stability CMS, which we have discussed in this article, 
may give the boundary curve of the pseudo gap region.

\end{document}